\documentclass[%
reprint,
superscriptaddress,
amsmath,amssymb,
aps,
prb,
twocolumn
]{revtex4-2}

\usepackage{amsmath,amssymb,amsfonts}
\usepackage[ruled,vlined,linesnumbered,resetcount,noend]{algorithm2e}
\usepackage{graphicx}
\usepackage{textcomp}
\usepackage{xcolor}
\usepackage{physics}
\usepackage{pifont}%
\newcommand{\xmark}{\ding{55}}%
\usepackage{graphicx}
\usepackage{dcolumn}
\usepackage{bm}
\usepackage{xcolor}
\usepackage{siunitx}

\usepackage[colorlinks]{hyperref}
\begin{document}

\title{Suppressing Correlated Noise in Quantum Computers via Context-Aware Compiling}

\author{Alireza Seif}%
\affiliation{%
	IBM Quantum, IBM T.J. Watson Research Center, Yorktown Heights, NY 10598, USA
}%
\author{Haoran Liao}%
\affiliation{%
	Department of Physics, University of California, Berkeley, CA 94720, USA
}%
\author{Vinay Tripathi}%
\affiliation{Department of Physics \& Astronomy, University of Southern California, Los Angeles, California 90089, USA}

\author{Kevin Krsulich}%
\affiliation{%
	IBM Quantum, IBM T.J. Watson Research Center, Yorktown Heights, NY 10598, USA
}%
\author{Moein~Malekakhlagh}%
\affiliation{%
	IBM Quantum, IBM T.J. Watson Research Center, Yorktown Heights, NY 10598, USA
}%
\author{Mirko Amico}%
\affiliation{%
	IBM Quantum, IBM T.J. Watson Research Center, Yorktown Heights, NY 10598, USA
}%
\author{Petar Jurcevic}%
\affiliation{%
	IBM Quantum, IBM T.J. Watson Research Center, Yorktown Heights, NY 10598, USA
}%
\author{Ali Javadi-Abhari}%
\affiliation{%
	IBM Quantum, IBM T.J. Watson Research Center, Yorktown Heights, NY 10598, USA
}%
\begin{abstract}
Coherent errors, and especially those that occur in correlation among a set of qubits, are detrimental for large-scale quantum computing. Correlations in noise can occur as a result of spatial and temporal configurations of instructions executing on the quantum processor. In this paper, we perform a detailed experimental characterization of many of these error sources, and theoretically connect them to the physics of superconducting qubits and gate operations. Equipped with this knowledge, we devise compiler strategies to suppress these errors using dynamical decoupling or error compensation into the rest of the circuit. Importantly, these strategies are successful when the context at each layer of computation is taken into account: how qubits are connected, what crosstalk terms exist on the device, and what gates or idle periods occur in that layer. Our context-aware compiler thus suppresses some dominant sources of error, making further error mitigation or error correction substantially less expensive. For example, our experiments show an increase of 18.5\% in layer fidelity for a candidate 10-qubit circuit layer compared to context-unaware suppression. Owing to the exponential nature of error mitigation, these improvements  due to error suppression translate to several orders of magnitude reduction of sampling overhead for a circuit consisting of a moderate number of layers. 
\end{abstract}
\maketitle


\section{Introduction}
\label{Introduction}

As quantum computers scale up, it becomes crucial to characterize, suppress, mitigate, and ultimately correct errors in the computation. {\em Error suppression} via better calibration or compilation is often the first line of defense, as it can prevent errors from surfacing with only small, constant overhead~\cite{viola1999dynamical,lidar1998decoherence, khodjasteh2009dynamically}. On the other hand, {\em error mitigation} has shown great success in removing errors from large computations, but involves additional circuit executions to improve the results, effectively trading (exponential) sample complexity for accuracy~\cite{PhysRevLett.119.180509,van2023probabilistic,tsubouchi2022universal}. 
Finally, {\em error correction} can robustly remove errors with polynomial or even constant circuit overhead, but this overhead can be large~\cite{shor1996fault, gottesman2013fault}. In both error mitigation and correction, the overhead can be significantly reduced if error suppression is deployed well.

Among the variety of possible errors, coherent quantum noise can harm computation quadratically more than incoherent (stochastic) errors. Therefore, their accurate characterization and suppression has become an important goal~\cite{kim2021hardware, gottesman2019maximally,kaufmann2023characterization}. Worse, some of these errors exhibit correlations across several qubits, thus spreading errors faster and making them particularly challenging to handle in error correction protocols~\cite{fowler2014quantifying, nickerson2019analysing}. In this paper, we present methods and experimental results for suppressing a wide range of correlated coherent errors, using the context of the hardware and the circuits that run on it. Throughout, we present a comprehensive characterization of several sources of errors, and discuss different strategies that suit each one.

In the near-term, the intersection of error suppression and error mitigation deserves a special attention. Error mitigation techniques typically use additional samples of the circuit and post-processing to improve the accuracy of the results.  In leading protocols such as Probabilistic Error Cancellation (PEC)~\cite{PhysRevLett.119.180509,van2023probabilistic} and Probabilistic Error Amplification (PEA)~\cite{kim2023evidence}, circuits are arranged in layers of gates that are subsequently twirled~\cite{knill2004faulttwirl,kern2005quantumtwirl,PhysRevLett.76.722twirl,geller2013efficienttwirl,wallman2016noisetwirl,PhysRevX.11.041039} by inserting random Pauli gates between them without changing the overall logic of the circuit. This twirling simplifies the structure of the noise in circuit layers and makes it amenable to error mitigation. The cost of error mitigation with PEC, as determined by its sampling overhead, grows exponentially with circuit size, and the base of the exponential is determined by the overall noise in the circuit. Therefore, even modest improvements in the errors can have significant impact on the total runtime of an error mitigated computation.  

The randomization introduced by twirling already suppresses the propagation of coherent errors in the circuit by converting those error channels to incoherent ones. However, at the same time the removal of the structure in the noise introduces challenges to its suppression. For instance, a known coherent error that could have been compensated by applying its inverse, can no longer be removed as simply. Given the extreme impact of reduced error rates on the sampling overhead, it is therefore crucial to suppress the known coherent errors in the circuit before twirling and converting those errors to incoherent ones.

An effective method to suppress coherent and temporally correlated single-qubit noise in circuits is Dynamical Decoupling (DD)~\cite{Hahn1950, Viola1999, Khodjasteh2005, Lidar-Brun:book, Jurcevic_Demonstration_2021, Ezzell_DD_2022}. However, going beyond the single-qubit case, and effectively inserting DD in a quantum circuit at scale is a challenging task. On one hand, the compiler must be aware of different crosstalk terms in the device Hamiltonian, noise spectrums on the qubits, and the physical gate calibrations used. On the other hand, the temporal and spatial structure of the circuit being executed plays an important role in determining the best sequence of DD gates to use. Our first contribution is an extensive characterization of correlated errors present in IBM quantum hardware, and a Context-Aware Dynamical Decoupling (CA-DD) framework that can dress arbitrary circuits with appropriate DD sequences. Our compiler works by coloring the qubits' interaction graph at each layer of the circuit based on the the contents of that layer and the underlying hardware.

Dynamical decoupling is very effective at suppressing known sources of crosstalk and temporally correlated noise. However, there are cases where it is difficult or undesirable to use DD. This could be the case, for example, when qubits are already actively participating in a gate and thus DD gates cannot be applied to them. Furthermore, gates used in DD sequences are themselves non-ideal and can introduce systematic errors or even crosstalk. Lastly, it may be challenging to fulfill the precise timing requirements of DD by the device controller, especially in cases of classical measurement and feed-forward operation, causing non-deterministic timing between some qubits while other qubits are idle.

To address this, we introduce a second compiler strategy termed Context-Aware Error Compensation (CA-EC) that compiles coherent errors directly into quantum algorithms, thus compensating for them without introducing extra overhead. The key insight enabling this is the fact that some errors are very well-characterized and remain constant over long periods of time, such as always-on $ZZ$ and Stark shift errors. This approach integrates well with the layered circuit structure of error mitigation protocols, where it is simple to categorize error patterns and find nearby gates that can be used for their compensation. Additionally, we show how simple dynamical decoupling sequences in conjunction with our error compensation method can achieve the performance of more complicated DD sequences.

We apply our compilation strategies to several quantum applications, including the simulation of 1-D Ising chain and a Heisenberg ring, the generic task of estimating circuit layer fidelity \cite{mckay2023layer}, and {circuits with intermediate measurements and feedforward}. In all cases, we demonstrate a reduction of errors, and consequently a decrease in error mitigation overhead. This presents a promising avenue for squeezing more performance from quantum computers, with almost no additional overhead. 

\section{Setup}
\subsection{Crosstalk and coherent errors in a quantum circuit}

Quantum crosstalk in a quantum processor refers to an intended quantum operation on a subset of qubits having unintended action on one or more qubits~\cite{Rudinger_Classifying_2018, Sarovar_Detecting_2020}. The nature and strength of crosstalk depends on the specific implementation of the quantum information processing device. Although our experiments are run on fixed-frequency cross-resonance (CR) processors \cite{Jurcevic_Demonstration_2021}, the errors we address are relevant to other platforms.  Broadly speaking, the crosstalk can be generated during single- and two-qubit gates, state-preparation and measurement, and idle times. Here, we focus on certain crosstalk-induced single- and two-qubit \textit{coherent} errors. 

A prevalent two-qubit crosstalk error in superconducting architectures is the always-on $ZZ$ interaction. This can originate from coupling to higher levels of the (transmon) qubits~\cite{Tripathi_Operation_2019, Magesan_Effective_2020, Malekakhlagh_First_2020} and is typically more problematic for fixed-frequency architectures. The Hamiltonian describing this error on a pair of qubits is 
\begin{equation}\label{eq:H11}
	H_{11} = \frac{\nu}{2} (-I\otimes Z - Z\otimes I + Z\otimes Z),
\end{equation}
where $\nu$ is the strength of the coupling that varies from pair to pair. This Hamiltonian indicates that qubits accumulate a phase when they are both in the excited state $\ket{1}$. Therefore, nearest-neighboring qubits experience an error of the form 
\begin{equation}\label{eq:U11}
	U_{11} = R_{zz}(\theta)\cdot[R_z(-\theta)\otimes R_z(-\theta)], 
\end{equation}
where $\theta = \nu \tau$, when they are idle for time $\tau$. 

When applying a gate, qubits are driven, which  alters this error. As we show, a major error that affects the spectator qubits, i.e., qubits in the vicinity of the qubits that the gate acts on, is a coherent $Z$ rotations generated by 
\begin{equation}\label{eq:hsingle}
	H_s = \frac{\nu}{2} Z, 
\end{equation}
which leads to error $R_z(\theta)$.  

In addition to these major sources of errors, spectators to both single-qubit or two-qubit gates can experience an AC Stark shift ($Z$ error) and potentially off-resonant $X/Y$-error~\cite{Malekakhlagh_Mitigating_2022} on the spectator~\cite{Wei_Characterizing_2023}. More generally, the degree by which the qubits are susceptible to various forms of crosstalk depends on proximity to frequency collisions~\cite{Malekakhlagh_First_2020, Hertzberg_Laser_2021, Zhang_High_2022, Heya_Floquet_2023}, and can be improved with post-fabrication laser annealing technique for fixed-frequency setups~\cite{Hertzberg_Laser_2021, Zhang_High_2022}.

Quantum crosstalk errors emerge in other quantum computing platforms as well. For example, in trapped ions, the crosstalk between the spectator qubits and the target in an application of a Mølmer-Sørensen gate \cite{Sorensen_quantum_1999} is a major source of error~\cite{PhysRevLett.129.240504}. In these systems, crosstalk occurs due to unwanted illumination of neighboring ions by laser during the gate operations. This leads to coherent errors of type $XX$ and $XY$, which were recently addressed and suppressed using dynamical decoupling techniques in Ref.~\cite{PhysRevLett.129.240504}. 

At the software level, these errors can be potentially suppressed after the compilation and scheduling of the circuit using dynamical decoupling (DD) (see e.g.~\cite{Lidar-Brun:book}) or the novel error compensation (EC) techniques that we introduce in this work. For both of these methods, well-characterized errors are essential to achieve the optimal performance. Moreover, crosstalk errors, as the definition suggests, depend not only on what happens to a qubit of interest, but also on other qubits in its vicinity. Therefore, the identification and characterization of the errors depend on the context and the operations in the circuit. This context dependence, however, provides an opportunity for the errors to be compensated by the operations preceding or following the error's occurrence. 

We discuss our two proposed error suppression techniques first in isolation, and then demonstrate how they can be modified and utilized depending on the context of the errors.  

\subsection{Dynamical decoupling} \label{sect:dd}
Dynamical decoupling (DD) is an open-loop control technique used for error suppression originating from coupling to unwanted interactions \cite{Lidar-Brun:book}. The basic motivation behind dynamical decoupling can be traced back to Hahn echo~\cite{Hahn1950} in nuclear magnetic resonance. These involve applying certain X gates in a specific order either periodically~\cite{Viola1998, Viola1999} or non-periodically~\cite{Uhrig2007, QDD2010} to qubits to suppress unwanted interactions. The central premise of DD is to apply a sequence of gates that effectively average the unwanted interaction Hamiltonian to zero. 

Consider a pure dephasing noise model where the unwanted interaction is given by $H_{\text{s}} = \frac{\nu}{2} Z $~\eqref{eq:hsingle}. Under such a model, the evolution of the system subjected to coherent phase noise for time $2\tau$ is characterized by the time-evolution unitary $U_{\text{s}} = R_z(2 \theta)$, where $\theta = \nu \tau$. If we now apply two $X$ pulses, at the beginning and in the middle of evolution,  the effective dynamics can be described by $U'_{\text{s}} = X R_z(\theta)X R_z(\theta)$. By noting that $X R_z(\theta)X = R_z(-\theta)$ we can simplify the expression for the effective dynamics to $U'_{\text{s}}= I$.

For stochastic colored noise, this cancellation is no longer exact, but such a DD sequence is still effective and can suppress the errors to order $\tau^2$. It is crucial to note that the above analysis presumes instantaneous pulses, which is an idealization. Nonetheless, DD remains a robust strategy for the suppression of temporally correlated incoherent errors induced by unwanted interactions.

\begin{figure}[t]
	\centering
	\includegraphics[width=\columnwidth]{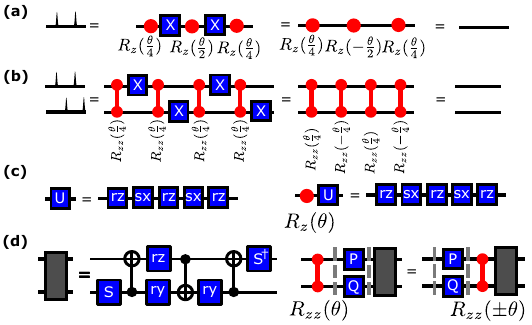}
	\caption{Decoupling and compensation of coherent errors.  (a)  A simple X2 dynamical decoupling pulse sequence can remove $Z$ errors by altering the sign of the accumulated phase during the evolution. (b) For two-qubit $ZZ$ errors, dynamical decoupling pulses have to be staggered to correctly alter the sign of the accumulated phase errors and remove them. (c) Coherent single qubit $Z$ errors can be removed by absorbing their inverse in the neighboring gates. Any single qubit gate can be decomposed into $3$ $R_z$ rotations, and two $\sqrt{X}$ gates. The absorption is simply achieved by modifying the first $R_z$ gate in this example and does not incur additional cost. (d)  The Cartan decomposition of $U_\text{can}$~\eqref{eq:canonical}. The $R_z$ rotation angle in the first qubit is $R_z(2\gamma-\pi/2)$. The two $R_y$ rotation angles in the second qubit are, from left to right, $R_y(\pi/2-2\alpha)$ and $R_y(2\beta-\pi/2)$. A two qubit $ZZ$ error can be compensated by an $R_{zz}$ rotation. In cases where a two-qubit gate of the $U_\text{can}$ form follows the error, it is possible to compensate the error with no overhead. However, moving the error past the twirling layer can change the sign of correction and has to be done carefully. }
	\label{fig:compiling_basics}
\end{figure}

While the above example illustrated a simple single-qubit DD sequence for suppressing single-qubit $Z$ errors, finding the sequence that effectively decouples a multi-qubit system with an arbitrary Hamiltonian to an arbitrary order in $\tau$ can be challenging~\cite{Khodjasteh2005}. 
In particular, to suppress a coherent two-qubit $ZZ$ error in addition to coherent single qubit $Z$ errors---i.e. those generated by $H_{11}$~\eqref{eq:H11}---it is necessary to apply DD pulses on two qubits in a staggered fashion (see Fig.~\ref{fig:compiling_basics}). This is because the $X\otimes X$ from DD commutes with the unwanted $Z\otimes Z$, and there will not be any cancellation. However, staggering the pulses ensures that there are periods of evolution accumulating opposite phases that eventually cancel and fully suppress the unwanted interactions~\cite{zhou2022quantum,shirizly2023dissipative}. 

While the implementation of DD in memory-like experiments, i.e., where the objective is to preserve a quantum state over an idling time, has been successful~\cite{zhou2022quantum,shirizly2023dissipative},  its integration into arbitrary quantum circuits requires meticulous attention to the context. Specifically, the presence of neighboring single or two-qubit gates necessitates careful alignments that maximize error suppression over all qubits. This becomes increasingly challenging as the complexity of the circuit scales. We formulate this problem as a constrained graph coloring one, and show that depending on crosstalk terms of the device and the presence of gate spectators at any given time, we may need $3$ or more colors even when the qubit graph is bipartite.

\subsection{Error compensation}
We introduce Error Compensation (EC) as a method to compensate coherent errors by absorbing the inverse of the errors in the gates preceding or following the error. The concept of EC is simple, however utilizing it in a computation requires attention to the context of the circuit. This method is especially useful when the compensation can be done with no overhead, i.e., the modification of the gates does not change the circuit runtime. Therefore, we need to identify cases where errors can be compensated and rules for moving the compensating operation throughout the circuit until they can be absorbed in an existing operation.  

Note that the major error processes that we discussed, i.e., $R_{z}(\theta)$ and $R_{zz}(\theta)$, can be simply inverted and compensated by applying $R_{z}(-\theta)$ and $R_{zz}(-\theta)$, respectively. The former is a single qubit rotation that is typically performed virtually in the controller. Therefore, generically, compensating it does not have an overhead. However, the $R_{zz}(-\theta)$ rotations require the application of two-qubit gates. Here, we discuss cases where there is a generic single-qubit gate or a two-qubit gate following the error. Then, in Sec.~\ref{sec:ca-ec} we discuss various strategies where these errors can be compensated by the compiler in general settings.

A single qubit gate from $\mathbb{SU}(2)$ can be implemented using the standard Euler angles and a basis of $R_z$ and $\sqrt{X}$ rotations.  Specifically, an arbitrary single qubit gate $U$ can be decomposed as 
\begin{equation}
	U = R_z(\alpha+\pi)\sqrt{X}R_z(\beta+\pi)\sqrt{X}R_z(\gamma), 
\end{equation}
where $\alpha,\beta,\gamma$ are the Euler angles. Now consider a single-qubit coherent error $R_z(\theta)$ that has occurred before $U$. It is straightforward to see that this error can be compensated by modifying $\alpha\to\alpha-\theta$. More general single qubit errors can be compensated similarly by adjusting the Euler angles. Many applications have circuits that consist of alternating layers of single-qubit and two-qubit gates. As we illustrate experimentally, in these cases, such errors can be compensated with no overhead. 

An arbitrary two-qubit gate $U$ in $\mathbb{SU}(4)$ can be implemented using 3 elementary CNOT gates as~\cite{PhysRevA.69.032315} (see also Fig.~\ref{fig:heisenberg}). In particular, operations of the form 
\begin{equation}\label{eq:canonical}
	U_{\text{can}} =\exp[i (\alpha X\otimes X+\beta Y\otimes Y + \gamma Z \otimes Z)]
\end{equation}
that appear in the simulations of spin models in Sec.~\ref{sec:application} can be implemented by the gate sequence depicted in Fig.~\ref{fig:compiling_basics}d. 
It is again straightforward to see that a $R_{zz}(\theta)$ error occurring before or after $U_{\text{can}}$ can be compensated by absorbing its inverse through modifying $\gamma$ to $\gamma-\theta/2$. In certain cases with layered circuits, there might be single qubit  gates between the $R_{zz}(\theta)$ error and the compensating gate. However, when those single qubit gates are Pauli gates, i.e. $I$, $X$, $Y$,  or $Z$, the compensation angle will simply change sign if $Z\otimes Z$ and the Pauli term $P\otimes Q$ do not commute (see Fig.~\ref{fig:compiling_basics}d).

\subsection{Experimental methodology}
{
	All the experiments in this work are performed on the IBM Quantum Platform~\cite{ibm}. 
	Except for the ideal expected results (labeled "Ideal" in Figs.~\ref{fig:ising} and Figs.~\ref{fig:heisenberg}), all the data points in the figures are obtained from experiments on quantum hardware, including those that characterize the noise and those that show noise suppression. Specific quantum systems used are mentioned in figure captions. The magnitude of coherent errors used for EC in this work are static and can be inferred from the reported backend information of IBM Quantum systems without the need for additional calibration~\cite{ibm}.} %
\section{Context-aware error suppression}

\subsection{Layered circuits and Pauli twirling}

Several quantum algorithms including those for quantum simulations (see e.g., Ref.~\cite{dalzell2023quantum} for a recent review) have circuits with a layered structure. Moreover, utilizing error mitigation technique necessitate arranging the circuit in a layered form to learn and mitigate the errors. In all these cases, it is crucial to suppress known sources of errors before attempting to correct or mitigate them to reduce the overhead. 

\begin{figure}
	\centering
	\includegraphics{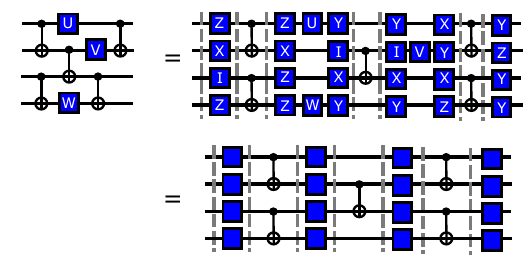}
	\caption{{An arbitrary circuit  can be stratified into layers of two-qubit and single-qubit gates. Pauli twirling adds single qubit Paulis surrounding the two qubit layers. The single-qubit gates (original and twirling components) are then recombined. The overall logical operation of the circuit remains unchanged.}}
	\label{fig:twirl}
\end{figure}

Here, we consider the problem of performing error suppression at scale. Specifically, we consider error suppression as a stage prior to error mitigation to reduce the mitigation overhead. In a typical workflow of error mitigation protocols such as PEC and PEA, an arbitrary circuit is stratified into alternating layers of single-qubit and two-qubit gates (see Fig.~\ref{fig:twirl}). Afterwards, the two-qubit gates are twirled, that is random single qubit gates are applied before and after the gates without altering the logical operation of the circuit. Twirling then simplifies the characterization and mitigation of errors~\cite{knill2004faulttwirl,kern2005quantumtwirl,PhysRevLett.76.722twirl,geller2013efficienttwirl,wallman2016noisetwirl,PhysRevX.11.041039}. Here, we focus on the specific case of Pauli twirling of Clifford two-qubit gate layers and assume that errors on single-qubit gates are gate-independent. Note that as any unitary operation can be realized by a gateset consisting of an entangling two-qubit gate and arbitrary single qubit rotations, any circuit can be stratified into alternating layers of arbitrary single qubit gates and Clifford two-qubit gates~\cite{barenco1995elementary, nielsen2010quantum}.

To realize a Pauli twirl, we apply random Pauli gates before and after the Clifford gate without altering the operations of the circuit. In this work we consider the two qubit gate layers to be composed of the hardware native Echoed Cross Resonance (ECR) gate~\cite{PhysRevA.93.060302, Sundaresan2020reducing, Jurcevic_Demonstration_2021, itoko2023three}, which is locally equivalent to a CNOT gate. Since the CNOT gate is a Clifford gate, it maps Pauli operators to other Pauli operators. Therefore, it is possible to find single qubit Pauli gates $P_1$, $P_2$ and $P'_1$ and $P'_2$ such that $(P_1 \otimes P_2)\cdot\text{CNOT}\cdot(P'_1 \otimes P'_2)$ = CNOT. This essentially realizes a Pauli twirl of the error channel of the gate (see Fig.~\ref{fig:twirl}). Such twirled error channels, are in a form known as a Pauli channel. Pauli channels with a local generator can be characterized and inverted efficiently~\cite{van2023probabilistic}. 

The random gates needed for Pauli twirling can be combined with the layer of single-qubit gates and implemented without additional overhead. This randomization can suppress the accumulation of coherent errors~\cite{cai2020mitigating}.  While this suppression can be useful if the errors are not known, it is important to remove the known errors whenever possible.  As mentioned earlier, in certain cases, known coherent errors can be absorbed in the preceding or following layers of gates. In the context of error mitigation, removing coherent errors before scrambling them can significantly improve the overhead.

\subsection{Context-dependent errors \label{sec:contexterr}}
Stratifying a circuit into layers introduces a structure that simplifies the classification and suppression of coherent errors in the layers based on the context. The alternating layer structure readily allows for overhead-free compensation of single-qubit coherent errors that only occur on idling qubits in layers of two-qubit gates by absorbing their inverses into the layers of single-qubit gates following or preceding them. The two-qubit coherent errors can be compensated with no overhead if there are generic two-qubit gates before or after them. Moreover, the structure imposed on the two-qubit gate layers simplifies the alignment of DD pulses to maximize error suppression. 

\begin{figure*}[t]
	\centering
	\includegraphics[width=\textwidth]{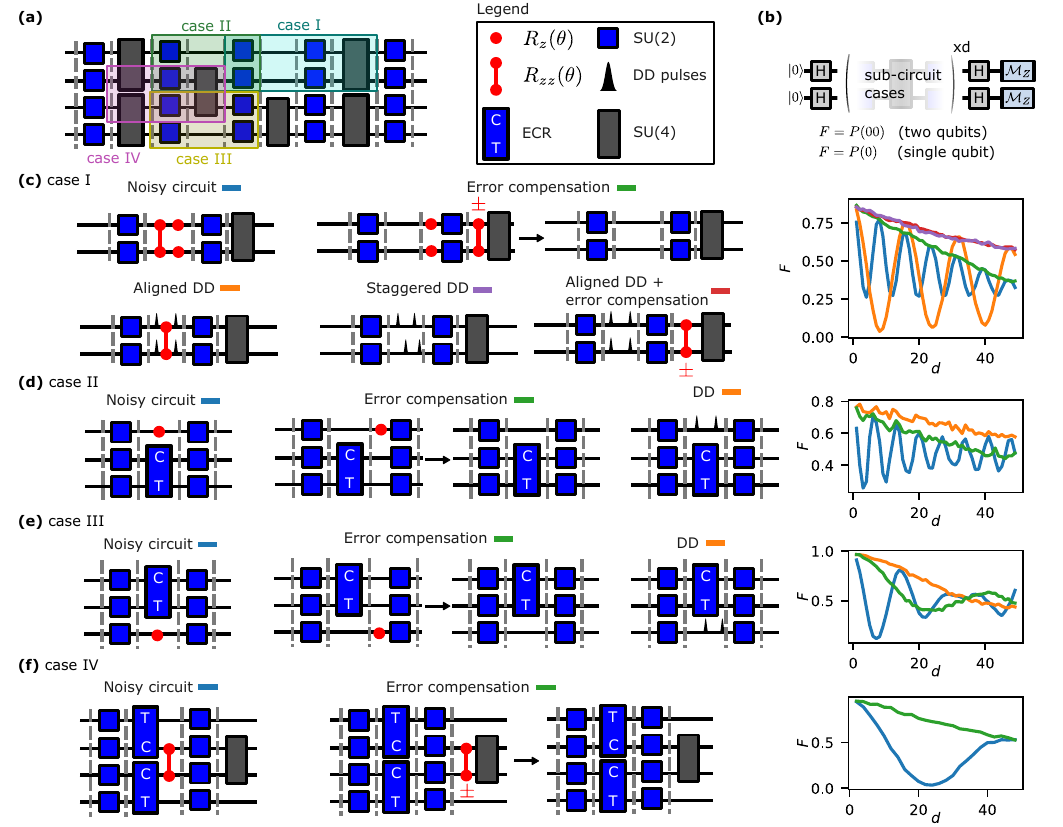}
	\caption{Context aware compiling and suppression of coherent errors using compensation or dynamical decoupling (DD). (a) Depiction of several cases where coherent errors can be identified and suppressed. (b) Ramsey experiments on one and two qubits are used to characterize the errors and their suppression. {The Ramsey fidelity $F$ should ideally stay at $F=1$ for all values of $d$. However, due to noise it deviates from the ideal value. Oscillations of $F$ in the subsequent panels are signatures of coherent noise.} (c) Idling neighboring qubits suffer from single-qubit $Z$ and two-qubit $ZZ$ errors. These errors can be removed by absorption into a two-qubit gate in the following layer,  by DD, or a combination thereof. Note that a conventional ``aligned'' DD on its own cannot fully remove these errors. The depth $d$ in the plot, refers to the number of idle intervals of $\tau = 500$ $\si{\nano\second}$. (d) Control spectator and (e) target spectator qubits both suffer from coherent $Z$ errors that can be absorbed into the single-qubit gate layers. (f) When the control qubits of two ECR gates are aligned, two-qubit $ZZ$ errors survive, but can be absorbed into a two-qubit gate in the following layer. In this case, DD cannot be applied without altering the gate operation.   {The colors and text above the circuits identify the corresponding curves in the fidelity plots.} The experimental data featured in the right column are from {\em ibm\_nazca} in panels (b-e) and from {\em ibm\_brisbane} in panel (f).}
	\label{fig:isolated_cases}
\end{figure*}

We consider several contexts in layers of two-qubit gates where coherent errors on idling qubits depend on the application and direction of the two-qubit gates on their neighbors (see Fig.~\ref{fig:isolated_cases}a)~\cite{PhysRevLett.117.210505}. We also examine the efficacy of various error suppression techniques by performing isolated Ramsey experiments. Specifically, we prepare the qubits, to which we apply error suppression, in the $|+\rangle$ state, let them evolve under noise for $d$ layers, apply error suppression, and measure the overlap of the final state with $|+\rangle$ state (see Fig.~\ref{fig:isolated_cases}a). Such measurements are sensitive to errors in the $Z$ basis and best illustrate the performance of our methods. Later we consider more general error metrics such as the layer fidelity in Sec.~\ref{sec:LF} and quantify the reduction in error mitigation overhead using our methods. 

The first example we consider (case I in Fig.~\ref{fig:isolated_cases}) is when there are two adjacent idle qubits for some time $\tau$. In this case, the qubits are affected by a coherent two-qubit error Hamiltonian $H_{11}$~\eqref{eq:H11}. Therefore, the total error is given by $U_{11}= R_{zz}(\theta)\cdot[R_z(-\theta)\otimes R_z(-\theta)]$~\eqref{eq:U11}, where $\theta=\nu \tau$. When these error are followed by a Pauli twirling layer and an arbitrary two-qubit gate, the $R_z$ terms can be compensated in the single qubit layer, and the $R_{zz}$ term can be moved past the twirling gates and compensated by the two-qubit gate that follows. Alternatively, context-aware staggered DD sequences can also suppress these errors~\cite{zhou2022quantum,shirizly2023dissipative}. However, if DD is applied on individual qubits without considering the context, the $X$ gates in DD align and cannot fully suppress the $U_{11}$ error. Specifically, while aligned DD cancels the $R_z(-\theta)$ terms, it cannot suppress $R_{zz}(\theta)$. In this case, one can again compensate the remaining error term in the following two-qubit gates. The results for all these suppression techniques for a varying idling time of $d\tau$ are experimentally demonstrated in Fig.~\ref{fig:isolated_cases}c. The remarkable improvement observed by error compensation validates our model for the major source of coherent noise in this experiment. The difference in the asymptotic decay rates of the error compensation and staggered DD curves is indicative of the presence of temporally correlated incoherent noise, which DD is effective at suppressing.

The second and third examples (cases II and III in Fig.~\ref{fig:isolated_cases}) involve spectators qubits of an ECR gate. The ECR gate has an echo pulse ($X$) on its control and rotary echo pulses on its target. These pulses act as DD sequences and cancel the $ZZ$ interaction between the control qubit and its spectator (Fig.~\ref{fig:isolated_cases}d), as well as the target qubit and its spectator (Fig.~\ref{fig:isolated_cases}e). Therefore, ignoring other sources of errors that will be discussed in the next Sec., the remaining errors are single-qubit $H_{Z} = - \frac{\nu}{2} Z$. This error can be compensated in single qubit gate layers that follow the ECR layer. To apply DD in these cases, it is important that the DD pulses on the spectators are correctly placed with respect to the echo and rotary echo pulses in the gate. To illustrate this point let $\tau_g$ denote the two-qubit gate time. For the control spectator we apply the sequence $\tau_g/4-X-\tau_g/2-X-\tau_g/4$ to ensure that the applied $X$ pulses form a staggered DD sequence with the control echo pulse. Note that in practice this timing has to be modified to account for the finite time required to apply the $X$ pulses.  For the target spectator, we apply the sequence $\tau_g/2-X-\tau_g/2-X$ to again ensure that the $X$ pulses form a staggered DD sequence with the target rotary echo pulses. These choices of alignments ensure that coherent errors between the qubits that the gate acts on and their spectators remain suppressed, while also suppressing  single qubit coherent errors on spectators. 

The fourth and last example we consider (case IV in Fig.~\ref{fig:isolated_cases}) is where the control qubits in two parallel ECR gates are adjacent. Here, the gate echo pulses on the control qubits align with each other and the $ZZ$ error reappears. In this case, DD cannot be applied and the only method to suppress the error is to compensate it using another two qubit gate. As shown in Fig.~\ref{fig:isolated_cases}f, the error compensation strategy is highly effective in suppressing these errors.

\subsection{Other sources of noise}

\begin{figure}[t]
	\centering
	\includegraphics[width=\columnwidth]{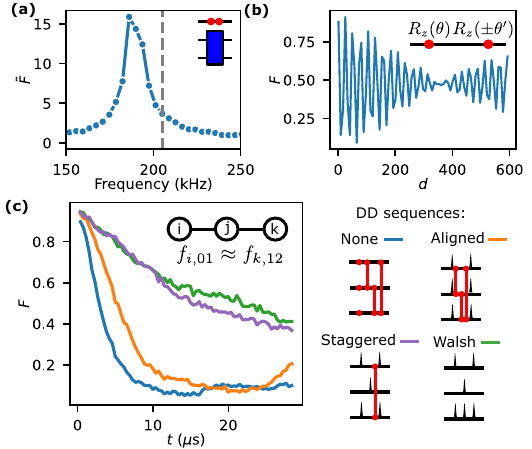}
	\caption{Characterization of some less dominant errors. (a) Pulses on neighboring qubits can induce coherent $Z$ errors on spectators. In this experiment on {\em ibm\_nazca}, we observe $20$ $\si{\kilo\hertz}$ of Stark shift (the distance between the peak and the frequency of always-on coupling indicated by the dashed line) due to the application of gates on neighboring qubits.  (b) Additional $Z$ errors can be caused by charge-parity fluctuations that results in a phase shift whose frequency $\delta$ changes very slowly in an experiment, but its sign switches from shot to shot. In the presence of another coherent phase evolution, the addition of these errors lead to beating in the oscillations as shown in this experiment on {\em ibm\_nazca}. (c)  Frequency collisions (inset) can cause next-nearest-neighbor (NNN) crosstalk. Progressively more cancellation will occur by going up the Walsh-Hadamard hierarchy to space gates on different qubits. Experimental data from {\em ibm\_sherbrooke} shows this suppression.}
	\label{fig:othersources}
\end{figure}
In addition to the dominant sources considered above,  there are other errors such as AC Stark shift, next-nearest neighbor $ZZ$ interaction, and slow $Z$ oscillations from charge-parity fluctuations~\cite{shirizly2023dissipative, catelani2011relaxation}. In this section we present the results of the characterization of these errors in experiments. However, since these errors are typically small compared to the ones considered in Section~\ref{sec:contexterr}, we do not actively seek to suppress them in the rest of the paper although the same methods could be applied.

A driven nonlinear quantum oscillator experiences a dynamic change in its effective frequency, referred to as the AC Stark shift. 
In weakly-anharmonic superconducting qubits such as transmons, the higher states can introduce non-negligible corrections. In cross-resonance architectures, a common instance is when applying ECR gate on a control qubit, or a single qubit gate on a qubit, leads to a non-negligible effective drive on its neighbors through the always-on qubit-qubit interactions. This drive spill-over can lead to Stark shift on spectator qubits in the form of $Z$ errors (see Fig.~\ref{fig:othersources}a). %

In addition to Stark shift, there are other $Z$ errors caused by charge-parity fluctuations through quasi-particle tunneling.  Such errors causes a phase shift whose frequency $\delta$ changes very slowly in an experiment, but its sign switches from shot to shot. That is, the single qubit error Hamiltonian with strength $\nu$, includes an additional $\pm \delta Z$ term 
\begin{equation}
	H = \frac{1}{2}(\nu \pm \delta) Z. 
\end{equation}
As the sign of the additional term is stochastic, error absorption techniques are not useful in this context, however, dynamical decoupling can still remove these errors. The strength of these errors vary in different systems and can be negligible in some cases. In Fig.~\ref{fig:othersources}b we show an example in experiments. Specifically, we perform a Ramsey experiment with a known rotation with frequency $\nu$. Therefore, the observed signal will have oscillations with frequencies $\nu \pm \delta$, which lead to beatings observed in Fig.~\ref{fig:othersources}b.

Finally, we consider $ZZ$ interaction between Next-Nearest-Neighboring (NNN) qubits. Such interactions are weak in cross-resonance architectures, typically of $\mathcal{O}(0.1$ $\si{\kilo\hertz})$, due to getting mediated by the middle qubit. Dependent on proximity to frequency collisions, however, we may run into qubit triplets in which such longer-range $ZZ$ interactions can be undesirably enhanced to even $\mathcal{O}(10\ \si{\kilo\hertz})$. One common scenario is the type-VI collision \cite{Hertzberg_Laser_2021} in which the $\ket{0}\rightarrow \ket{1}$ transition of one qubit is approximately resonant with the $\ket{1} \rightarrow \ket{2}$ transition of the NNN qubit. %

In such scenarios with all-to-all $ZZ$ interaction, the idea of staggered DD can be generalized using the more advanced Walsh-Hadamard DD sequence~\cite{Leung_Simulation_2002,pawel2006}. This technique is based on shifting the $X$ pulses on each qubit in time, such that the net accumulation of the $+$ and $-$ sign $Z$ error on each qubit due to any mutual $ZZ$ interaction is canceled out. Such sequences can be constructed using a sign matrix with number of rows and columns equal to the number of qubits and layers, where $ZZ$ cancellation translates into zero inner product between any two rows \cite{Leung_Simulation_2002, Lidar-Brun:book}. To achieve this for three qubits, we can implement the sequence shown in Fig.~\ref{fig:othersources}c.

Table~\ref{tab:example} summarizes the different errors and suppression techniques that we have covered. Next, we describe how a compiler can address these errors for arbitrary circuits.
\begin{table}[ht]
	\centering
	\begin{tabular}{ c c c c}
		\textbf{Error} & \textbf{Source} & \multicolumn{2}{c}{\textbf{Suppression}} \\
		\cline{3-4}
		& & \textbf{EC} & \textbf{DD} \\
		\hline
		$Z$ (idle) & Always-on & Phase shift & Any \\
		$ZZ$ (idle) & Always-on & Absorb & Staggered  \\
		$ZZ$ (active) & Always-on & Commute/absorb & \xmark \\
		Stark $Z$ & neighboring gate & Phase shift & Any \\
		Slow $Z$ & Quasi-particles & \xmark & Any \\
		NNN $ZZ$ & Freq. collisions & \xmark & Walsh \\
	\end{tabular}
	\caption{The coherent errors characterized experimentally in this paper, and how each is suppressed.}{ Active qubits are those that participate in a gate (here specifically control). Experimental characterization of errors in rows 1-3 are shown in Fig.~\ref{fig:isolated_cases}. The less frequent errors mentioned in rows 4-6 are characterized in Fig.~\ref{fig:othersources}.}
	\label{tab:example}
\end{table}
\section{Compiler design}

Having established different sources of correlated errors and how to suppress them in small experiments, we now turn to the design of a compiler to automatically implement these suppression techniques for arbitrary circuits. The first compilation strategy identifies periods in which neighboring qubits are jointly idling, and resolves via graph coloring a suppression sequence to be inserted based on the device Hamiltonian, connectivity of idling qubits, and gates being executed on neighboring qubits (the circuit context). The second automates the identification of locations in the circuit where coherent errors are known to accumulate, again from the circuit context, and inserts error-compensating gates to refocus the undesired evolution. {Both suppression methods scale favorably with both the circuit and device size, with complexity of $\mathcal{O}(d^2n)$ for CA-DD and $\mathcal{O}(dn)$ for CA-EC with circuit depth $d$ and number of device qubits $n$.}

\subsection{Context-aware dynamical decoupling (CA-DD)}

This algorithm proceeds in four phases, shown in Algorithm~\ref{alg:ca-dd}. First, a qubit crosstalk graph is built based on knowledge of the device Hamiltonian from calibration and crosstalk characterization data. Often, this means having an edge between neighboring qubits, but in collision conditions there may be additional edges connecting next-nearest neighbors (e.g. as characterized in Fig.~\ref{fig:othersources}c). Next, the layers of an input circuit are scanned to identify qubit idle periods which may be candidates for dynamical decoupling, depending on their duration and the characteristics of their respective qubits. These periods are greedily collected into groups of delays which overlap in time and are adjacent on the crosstalk graph. 

The algorithm then evaluates each group recursively, examining at the entry and exit of each variable-width delay instruction to identify the longest time period and largest collection of qubits which are candidates for dynamical decoupling based on the duration, crosstalk interaction, and number of jointly idling qubits. The remainder of the group is split and each residual is evaluated in the same way. These delay groups are then analyzed within their circuit and device context to select an optimal suppression sequence. This includes examining the circuit context to account for concurrent gates on qubits adjacent to the idling group, as described in Sec.~\ref{sec:contexterr},
such that adjacent sequences do not negate the effect of one another. 

\begin{figure*}[ht!]
	\centering
	\includegraphics[width=0.9\linewidth]{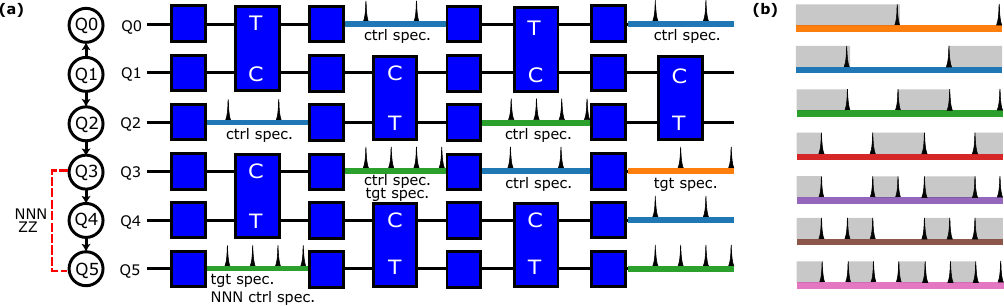}
	\caption{Context-aware dynamical decoupling. a) An example qubit connectivity, upon which a crosstalk graph is built (note the existence of a next-nearest-neighbor crosstalk). A $4$-layer circuit is shown (interleaved with twirls). In each layer, the idle qubits are colored according to Algorithm~\ref{alg:ca-dd}. Based on these colors, Walsh-Hadamard sequences are inserted to suppress correlated $Z$ and 
		$ZZ$ (and $ZZZ$) errors within that context. b) The location of pulses in the first $7$ Walsh-Hadamard sequences. Each sequence corresponds to a different color after graph node coloring. Each sequence suppresses $Z$ (as evidenced by the balanced shaded area in each row). In addition, $ZZ$ is suppressed between any two of these sequences (as evidenced by the balanced parity of the shaded areas in each pair of rows). The compiler aims to use the lowest number of pulses (minimize colors subject to constraints).}.
	\label{fig:ca-dd}
\end{figure*}

A greedy graph coloring algorithm is employed, as shown in Fig.~\ref{fig:ca-dd}. The spectator qubits to gates provide the initial constraints in the graph coloring problem. For each layer in the circuit, start by coloring the qubits participating in ECR: orange for control and blue for target. This ensures two things: first, that the control spectator is not colored orange, and thus its echos will be staggered with that of the control itself. Second, that the target spectator is not colored blue, and thus we get $Z$ suppression on this spectator without undoing the effect of rotary pulses at canceling $ZZ$ between the target and the spectator during those intervals. From there, we color the rest of the idle nodes via a greedy assignment beginning with those already constrained by the coloring of adjacent ECR gates, ensuring that no two neighbors in the crosstalk graph are colored the same. Whenever there is a conflict and the desired color cannot be applied, we can use the next level of the Walsh-Hadamard hierarchy, as initially described in Sec.~\ref{sec:contexterr} and further illustrated in Fig.~\ref{fig:ca-dd}. This heuristically minimizes the number of DD pulses by staying in lower levels of the Walsh hierarchy, yet ensure that adequate crosstalk suppression is achieved on all pairs. {Note that with a finite number of colors, constructing higher order Walsh sequences has a constant cost, and we can use a dictionary of pre-built sequences (Fig.~\ref{alg:ca-ec}b) in the compiler.}

\begin{algorithm}[t]
	\caption{Context-aware dynamical decoupling}\label{alg:ca-dd}
	
	\SetAlgoLined
	
	\KwIn{Scheduled circuit $S$,
		device description $\mathcal{H}$,
		minimum duration to suppress $D_\text{min}$,
		{dictionary} of dynamical decoupling sequences $L_\text{DD}$,
	}
	\KwOut{Error-suppressed circuit $S'$}
	
	\SetKwData{SQI}{$s$}
	\SetKwData{AIGS}{$gs$}
	\SetKwData{AIG}{$g$}
	\SetKwData{OQI}{idle\_single}
	\SetKwData{GI}{grouped\_idle}
	\SetKwData{WIDEST}{$w$}
	\SetKwData{ISS}{delay\_{interval}}
	\SetKwData{BEFORE}{before}
	\SetKwData{AFTER}{after}
	\SetKwData{COLORING}{coloring}
	
	\SetKwFunction{Sort}{Sort}
	\SetKwFunction{Append}{Append}
	\SetKwFunction{Break}{Break}
	\SetKwFunction{Pop}{Pop}
	\SetKwFunction{QubitsAdjacent}{QubitsAdjacentOnDevice}
	\SetKwFunction{DelaysOverlap}{DelayIntervalsOverlap}
	\SetKwFunction{ColorGraph}{ColorGraph}
	\SetKwFunction{GreedyColor}{GreedyColor}
	\SetKwFunction{CollectJointDelays}{CollectJointDelays}
	\SetKwFunction{BuildInteractionGraph}{BuildInteractionGraph}
	\SetKwFunction{InsertDDSequenceByColor}{ApplyDDSeqByColor}

	\SetKwProg{Fn}{Function}{:}{}
	\SetKw{Return}{return}
	
	\BlankLine
	\Begin{
		$G \leftarrow$ \BuildInteractionGraph{$\mathcal{H}$}\;
		$I \leftarrow$ \CollectJointDelays{$S$, $G$, $D_\text{min}$}\;
		$C \leftarrow$ \ColorGraph{$I$, $G$, $S$, $\mathcal{H}$}\;
		$S' \leftarrow$ \InsertDDSequenceByColor{$S$, $C$, $L_\text{DD}$}\;
		
	}
	\Fn{\CollectJointDelays{$S$, $G$, $D_\text{min}$}}{
		\SQI $\leftarrow$ [\textnormal{single qubit \emph{delay} instructions in} $S$ with duration $\geq D_\text{min}$]\;
		\AIGS $\leftarrow$ [\textnormal{greedy collection of delays {into groups} adjacent on device and overlapping in time}]\;
		$I \leftarrow \emptyset$\;
		\While{\AIGS $\neq \emptyset$}{
			\AIG $\leftarrow$ \AIGS.pop(); \WIDEST $\leftarrow \emptyset$\;
			\For{\ISS $\in$ \AIG}{
				$m \leftarrow \ISS.num\_{idling}\_qubits$\;
				\If{$m >$ \WIDEST.num\_{idling}\_qubits}{
					\WIDEST $\leftarrow$ \ISS\;  %
				}
			}
			\BEFORE, \WIDEST, \AFTER $\leftarrow$ \AIG.split\_at(\WIDEST)\;
			\AIGS.append(\BEFORE, \AFTER)\;
			$I$.append(\WIDEST)\;
		}
		\Return{$I$}\;
	}
	\BlankLine
	\Fn{\ColorGraph{$I$, $G$, $S$, $\mathcal{H}$}}{
		$C \leftarrow \emptyset$\;
		\For{\ISS $\in I$}{
			$c\leftarrow \emptyset$\;
			\For{gate $\in$ S.adjacent\_gates(\ISS)}{
				{
					\If{gate.type == ECR}{
						$c$.assign(gate.control, ``orange'')\;
						$c$.assign(gate.target, ``blue'')\;
					}
				}
			}
			\GreedyColor($c$, {\ISS, $G$})\;
			$C$.append($c$)\;   
		}
		\Return{$C$}\;
	}
	
\end{algorithm}

\subsection{Context-aware error compensation (CA-EC)\label{sec:ca-ec}}

The second technique employed by our compiler is the automated identification, accumulation, and compensation of coherent crosstalk errors. Similar to the approach employed above, an input circuit is scanned to identify periods under which, from static crosstalk terms in the device Hamiltonian, coherent noise will be known to accumulate. Circuit instructions which invert the effect of these noise sources are inserted before or following the period where they will occur. They are subsequently commuted through existing circuit instructions (or anti-commuted, tracking any needed corrections) to a location in the circuit where they are either synthesized into circuit gates directly or absorbed as corrections into existing circuit blocks. This is described in Algorithm~\ref{alg:ca-ec}.

First, we initialize a dictionary mapping each qubit and each pair of qubits in the connectivity graph to the final error compensation accumulated on those qubits. This is initially $Z(0)$ for individual qubits and $ZZ(0)$ for pairs of qubits. Next, we scan over the scheduled circuit layer by layer, and identify how correlated errors affect each of the entries in our map. This is done in accordance with the isolated cases outlined in Fig.~\ref{fig:isolated_cases}. Taking into account the $ZZ$ rates and the duration of each layer, we calculate compensation angles affecting each entry in the dictionary and add to the ones accumulated so far. When we carry the compensation dictionary to the next layer, we may pass through a twirl layer. Depending on whether the twirl Paulis commute or anti-commute with each of the $Z$ or $ZZ$ compensations, we may need to keep the compensation angle's sign, or flip it. 

If we reach a layer where an entry in the compensation dictionary can no longer be commuted through, we look at the succeeding gate (and more gates downstream if supported on the same qubits) to see if the compensation can be absorbed into those existing gates. This is the case, for example, when the succeeding gate is itself a $ZZ$ rotation (as in QAOA or Ising model applications) or a more general Heisenberg or $\mathbb{SU}(4)$ gate, in which case a $ZZ$ can be absorbed into it at no additional cost. Otherwise, we explicitly insert the compensation as a gate into the circuit. We reset the compensation for this entry of the dictionary, and proceed to the next layer. We repeat until all scheduled circuit layers are exhausted.

To avoid introducing extra errors from the compensating gates themselves (when they cannot be absorbed), a final optimization is performed by our compiler. In these cases, observing that the compensating angle is often small, we use pulse stretching to natively implement $ZZ(\theta)$~\cite{gokhale2020optimized, stenger2021simulating}, which saves substantially on gate error compared to implementing it from two CNOTs due to having a much shorter pulse. This is irrelevant for $Z$ compensations, since those are implemented virtually via microwave phase shifts and have zero cost~\cite{mckay2017efficient}.

\begin{algorithm}[t]
	\caption{Context-aware error compensation}\label{alg:ca-ec}
	\SetAlgoLined
	
	\KwIn{Scheduled circuit $S$, connectivity graph $C$,
		characterized $ZZ$ crosstalk rates $R$
	}
	\KwOut{Error-suppressed circuit}
	
	\SetKwIF{If}{ElseIf}{Else}{if}{then}{else if}{else}{endif}
	\SetKwFunction{Continue}{Continue}
	
	$compensation\_1q \leftarrow \{q: \emptyset \textnormal{ for n in }C.nodes()\}$\;
	$compensation\_2q \leftarrow \{e: \emptyset \textnormal{ for e in }C.edges()\}$;
	
	\tcc{{Split circuit into layers with either only 1q or only 2q gates}}
	\For{layer $\in$ Layers($S$)}{
		\If{layer.is\_{2}q}{
			\For{$q \in compensation\_1q$}{
				\For{$p \in C.neighbors(q)$}{
					\If{$layer[p, q]$ refocuses $ZZ$}{
						\Continue\;
					}
					\Else{
						$\nu \leftarrow R[p,q]$\;
						$\theta \leftarrow 2\pi * \frac{\nu}{2} * \tau$\;
						$compensation\_1q[q]$ += $\theta$\;
						$compensation\_1q[p]$ += $\theta$\;
					}
				}
			}
			\For{$edge \in compensation\_2q$}{
				\If{$layer[edge]$ refocuses $ZZ$}{
					\Continue\;
				}
				\Else{
					$\nu \leftarrow R[edge]$\;
					$\theta \leftarrow -2\pi * \frac{\nu}{2} * \tau$\;
					$compensation\_2q[edge]$ += $\theta$\;
				}
			}
		}
		\Else{
			\For{$({compensation}, error) \in (compensation\_1q, compensation\_2q), (Z, ZZ)$}{
				\tcc{{Update compensation if error commutes or anti-commutes with layer}}
				\For{{$key \in compensation$}}{
					
					\If{$\{layer[key], error\} == 0$}{
						$compensation[key]$ *= -1\;
					}
					\ElseIf{$[layer[key], error] == 0$}{
						$compensation[key]$ *= 1\;
					}
					\Else(\tcc*[h]{{Insert correction}}){
						${layer}$.insert($compensation[key]$)\;
						$compensation[key] \leftarrow \emptyset$\;
					}
				}
			}
		}
	}

\end{algorithm} 

Our context-aware dynamical decoupling and error compensation techniques are compatible. The latter is more effective in refocusing coherent noise that occurs during gate applications and thus cannot admit the insertion of additional decoupling gate sequences, and the former is more general in that it can suppress coherent and incoherent noise sources. {We explore the combination of these methods in the next section. } %
\section{Applications}
\label{sec:application}
We now study several interesting circuits as benchmarks. {The qubit layouts in these applications are selected to include error mechanisms depicted in Fig.~\ref{fig:isolated_cases} to highlight the detrimental effect of coherent noise.}

\subsection{Ising model}
\begin{figure}[t]
	\centering
	\includegraphics[width=\columnwidth]{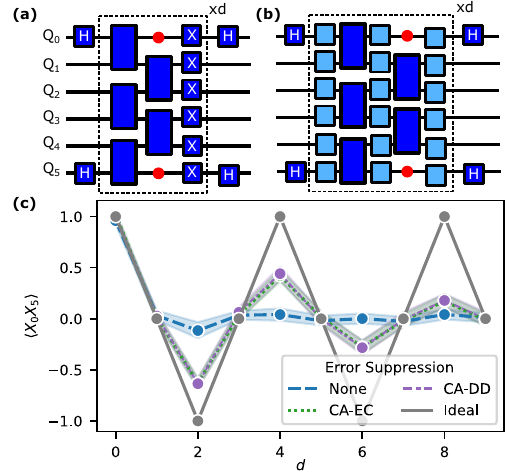}
	\caption{Floquet Ising evolution with error compensation. (a) The circuit for the Floquet Ising model consists of $d$ steps, each with two layers of two-qubit gates, and a layer of single qubit gate. (b) The two-qubit gate layers are twirled. The idling period in one of the layers results in additional $Z$ errors at the boundary. (c) The experimental results from {\em ibm\_nazca} show that compensation of errors (CA-EC) or {decoupling them (CA-DD)} at the boundary (red circles in panels (a) and (b))  significantly improves the results compared to the case where layers are only twirled without error suppression.}
	\label{fig:ising}
\end{figure}

We consider the Floquet time-evolution of an Ising-type model at the Clifford point, similar to those studied in Ref.~\cite{kim2023evidence}. Specifically, each Floquet step consists of a layer of ECR on even-odd qubits, and a layer of ECR on odd-even qubits, followed by a layer of single qubit $X$ gates. This serve as a test case,  where boundary qubits (Q0 and Q5 in Fig.~\ref{fig:ising}) are initialized in $\ket{++}$ while the other qubits are in $\ket{0}$. The expectation values $XX$ for spins at boundaries, i.e., $\langle X_0 X_5\rangle$, oscillates between $-1$ and $+1$. We recover this behavior through our CA-EC {and CA-DD techniques}, as shown in Fig.~\ref{fig:ising}.

\subsection{Heisenberg model}
\begin{figure}[ht!]
	\centering
	\includegraphics[width=\columnwidth]{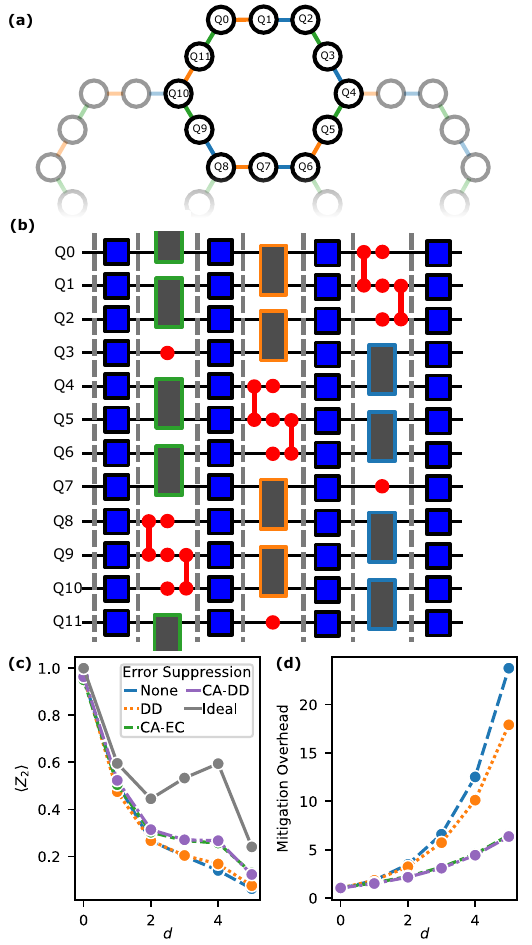}
	\caption{Heisenberg model. (a) A ring on a heavy hex architecture, where the interaction color indicates layers of two qubit gates inside a time step.  (b) A mapping of the Heisenberg ring onto a 12-qubit circuit with a periodic boundary condition. The schematic shows one time step. The edge color of the two-qubit interaction corresponds to those shown in (a). The grey gates are $U_{\text{can}}$~\eqref{eq:canonical}. Red indicates $ZZ$ and $Z$ errors on the qubits, most of which will be compensated into previous/subsequent Heisenberg interactions. (c) Performance of context-aware error suppression in a Heisenberg experiment on {\em ibm\_nazca}. {(d) Estimated error mitigation overhead for error suppression techniques in panel (c) shows that CA-EC and DD reduce the overhead by more than $3.5\times$ and $2.75\times$ over no suppression and DD, respectively.}}.
	\label{fig:heisenberg}
\end{figure}

We demonstrate our context-aware compiling for the task of generating first-order Trotterized dynamics of the Heisenberg model with a periodic boundary condition in the absence of external field, on $12$ spins. The Heisenberg model is a cornerstone in quantum magnetism, providing a framework for understanding spin interactions in condensed matter systems, particularly in exploring phenomena such as quantum phase transitions and criticality~\cite{sachdev_2011}. %
The Heisenberg Hamiltonian of an $N$-spin chain in the absence of external field can be written as
\begin{equation}
	H=-\frac{1}{2}\sum_{\langle i,j\rangle}^N\left(J_x X_i X_{j}+J_y Y_i Y_{j}+J_z Z_i Z_{j}\right),
\end{equation}
where $J_x$, $J_y$, $J_z$ are the coupling constants and $\langle i,j\rangle$ indicates that the sum is performed on adjacent qubits on the lattice. The unitary operator generated by this Hamiltonian, $\exp(-iHt)$, can be decomposed in terms of the canonical two-qubit gate $U_{\text{can}}$~\eqref{eq:canonical} as shown in Fig.~\ref{fig:heisenberg}b, where the angles $\alpha,\beta,\gamma$ are defined as $-J_i t/2,\ i\in\{x, y, z\}$, respectively. 
In a heavy hex lattice, implementing the discretized time-evolution of the Heisenberg model requires $3$ layers of two-qubit unitaries per time step, as shown with three colors in Fig.~\ref{fig:heisenberg}a,c. Here we consider this evolution in a ring of 12-qubits. 

On this ring, neighboring qubits $i$ and $j$ are exposed to $ZZ$ coherent noise during any idling periods in each time step, whose error Hamiltonian is $H_{11}=\nu_{i,j}/2(-Z_i-Z_{j}+Z_i Z_{j})$~\eqref{eq:H11}, where $\nu_{i,j}$ is the strength of the coherent errors. As mentioned in Sec.~\ref{sec:contexterr}, such errors accumulated over an idling period of $\tau$ can be corrected by a compensating unitary, i.e., $U_{\text{comp}}= R_{zz}(-\theta)\cdot[R_z(\theta)\otimes R_z(\theta)]$ where $\theta=\nu\tau$. 

Our context-aware compiling then absorbs the $ZZ$ part of the correction into a time step of Heisenberg interaction. We remark that since $XX$, $YY$, and $ZZ$ all commute with each other in a Heisenberg interaction, we can absorb the correction unitary into either the previous or the following interaction unitary. 
For instance, in Fig.~\ref{fig:heisenberg}, the compensating unitary $U_{\text{comp}}$ for the idling period in the second two-qubit gate layer on Q4 and Q5 can be absorbed into the $ZZ$ Heisenberg interaction in the first two-qubit gate layer, whereas the compensating unitary for the idling period in the second two-qubit gate layer on Q5 and Q6 can be absorbed into the $ZZ$ Heisenberg interaction in the third layer. 

If we Pauli twirl the entire circuit, there will be two identical pairs of Pauli gates conjugating every two-qubit idling period. To absorb the coherent $R_{ZZ}(\theta)$ errors accumulating over any idling period, we need to push them through the twirling gates before absorbing them into the $ZZ$ interaction in the Heisenberg interaction. As explained in Sec.~\ref{sec:contexterr}, if the Pauli twirling gates do not commute with $ZZ$, the accumulated $ZZ$ error evolution acquires a negative sign (see Fig.~\ref{fig:compiling_basics}). The compensating unitary for the single qubit $Z$ errors, i.e., $R_z(\theta)$, can easily be absorbed into any neighboring single qubit unitary, e.g., from the twirling or the Heisenberg interaction decomposition (see Fig.~\ref{fig:compiling_basics}). 

We experimentally compare the performance of several error suppression techniques discussed in this work, namely CA-EC, CA-DD, and regular DD, in the Heisenberg ring. 
As shown in Fig.~\ref{fig:heisenberg}c,  we observe that with no error suppression (except readout correction and Pauli twirling), the distinguishing features of the dynamics of a generic single qubit observable $\langle Z_2 \rangle$ are not observable. By adding CA-EC or CA-DD the features of the dynamics such as oscillations at $d=4$ are recovered. However, not considering the context and applying DD individually to qubits does not noticeably improve the results. Note that this is a fairly large circuit, consisting of $180$ CNOTs and a CNOT-depth of $45$. Having boosted the ``raw'' signal, it is now possible to deploy error mitigation methods to recover the rest of the signal at substantially reduced overheads. {In fact, using a simple global depolarization error model we can estimate this overhead (see e.g., Ref.~\cite{emreview2023}). Specifically, we scale expectation values at depth $d$ by $A \lambda^d$ such that they are as close to ideal values as possible. Here, $A$ and $\lambda$ capture readout and layer errors, respectively. Since scaling the signal modifies its variance, the sampling overhead can be obtained by the ratio of the scaling factors squared. The results shown in Fig.~\ref{fig:heisenberg}d indicate that CA-DD and CA-EC improve the error mitigation overhead by more than $3.5\times$ and $2.75\times$ over no suppression and DD, respectively.}

\subsection{Layer Fidelity}
\label{sec:LF} 

Lastly, we consider a very generic form of computation, that is an arbitrary layer of simultaneous two-qubit gates applied to some qubits on the device. Any circuit can be translated to layers of this form, as is routinely done in PEC or PEA error mitigation. The fidelity of these layers plays a direct role in error mitigation overhead, with overhead becoming exponentially better as layer fidelities improve.  %
{We evaluate the layer fidelity following a procedure similar to that described in Ref.\cite{mckay2023layer}. First, we partition the qubits into disjoint groups of pairs subjected to gate operations, adjacent idle pairs, and individual idle qubits. We then simultaneously measure the average fidelity within each partition by preparing and measuring the fidelity in the corresponding full two-qubit or one-qubit Pauli basis. This involves standard steps of preparing the circuit in a Pauli basis using 1-qubit Clifford gates, applying randomly-twirled versions of the layer to diagonalize the noise in the Pauli basis, and iteratively applying it for a depth of $d$ to amplify the error. We revert the computation by applying another Pauli basis change so that ideally we would implement the identity operation~\cite{erhard2019characterizing,van2023probabilistic,mckay2023layer}. Additionally, we incorporate a twirling layer before readouts, which  diagonalizes the readout errors through averaging over systematic errors~\cite{van2022model}. From the decay rate of the circuit fidelity as a function of $d$ we obtain the fidelity in each partition. Finally, we determine the layer fidelity using the product of the disjoint fidelities~\cite{mckay2023layer}}

\begin{figure}[t]
	\centering
	\includegraphics[width=\columnwidth]{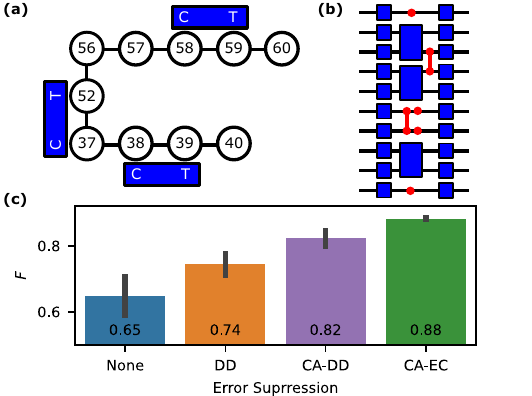}
	\caption{ {Benchmarking layer fidelities. (a) The layout for a 10-qubit layer on {\em ibm\_nazca}, with 3 ECR gates and $4$ idle qubits. (b) The twirled layer that we benchmark with its coherent errors highlighted in red. (c) We see a $1.26\times$ ($1.36\times$) improvement in layer fidelity by employing CA-DD (CA-EC), compared to $1.14\times$ with context-unaware DD. %
			CA-EC performs  better in this layer than CA-DD, possibly due to the presence of $ZZ$ between Ctrl-Ctrl on Q37 and Q38, which DD cannot suppress}}
	\label{fig:lf}
\end{figure}

To compare the various methods presented in this paper, we benchmark the layer depicted in Fig.~\ref{fig:lf}. {This sparse layer contains adjacent idle qubits and adjacent control qubits that are especially susceptible to correlated coherent noise. While the former can be suppressed by CA-DD the latter can only benefit from CA-EC.}  Without any error suppression (except for readout) we observe a circuit layer fidelity of {$\mathcal{LF}_\text{bare} = 0.648 \pm{0.058}$}. This fidelity increases to {$\mathcal{LF}_\text{DD}=0.743 \pm{0.032}$} using DD to {$\mathcal{LF}_\text{CA-DD}=0.822\pm{0.024}$} applying context-aware DD and to {$\mathcal{LF}_\text{CA-EC}=0.881\pm{0.002}$} applying error compensation. These increases in layer fidelity translate to reduced factors from {$\gamma_\text{bare} \approx 2.38$} to {$\gamma_\text{DD} \approx 1.81$, $\gamma_\text{CA-DD} \approx 1.48$} and to {$\gamma_\text{CA-EC} \approx 1.29 $, where $\gamma$ determines the error mitigation  sampling overhead for the same layer \cite{van2023probabilistic, mckay2023layer}}. To put it into perspective, the reduction in sampling overhead for mitigating a single layer compared between ordinary DD / CA-DD and ordinary DD / CA-EC is a factor of {$1.2\times$ and $1.4\times$,} respectively. However, the sampling overhead scales exponentially in depth, i.e. for a circuit of $10$ layers, the reduction is {$\sim 7\times$ and $\sim 30\times$}.%

\subsection{{Dynamic circuits}}
\begin{figure}[t]
	\centering
	\includegraphics[width=\columnwidth]{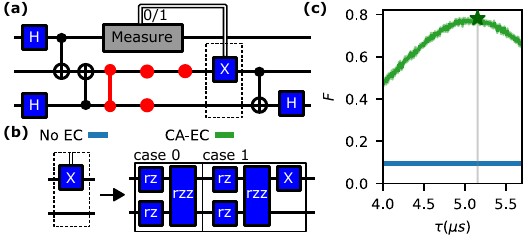}
	\caption{{Error compensation for dynamic circuits. (a) Quantum circuit for preparing a Bell state using mid-circuit measurement on an auxiliary qubit (top) and a conditional $X$ gate on one of the data qubits (middle).  The $ZZ$ and  $Z$ errors during measurement and feedforward are shown in red. In the last stage, we measure the Bell fidelity ($F$) by applying a  CNOT and a Hadamard (H) gate. This simplifies the computation $F$ to evaluating the probability of the data qubits to return to $|00\rangle$. (b) The errors are compensated by appending the inverse of the accumulated $ZZ$ and $Z$ rotations to the conditional operation in (a). When the intermediate measurement results in $1$ (case 1), there would be an additional $Z$ rotation on the middle qubit. (c) Fidelity $F$ with varying  estimate of the idle time $\tau$. The optimal $\tau$ corresponds to the best estimate of the measurement and feedforward time in the experiment. With error compensation, $F$ is increased by more than $8\times$. Experiments are performed on {\em ibm\_nazca}.}}
	\label{fig:dyncir}
\end{figure}

{
	Dynamic circuits have recently emerged as a promising technology with implementations in different architecture~\cite{racetrack2023,baumer2023efficient}. These circuits involve mid-circuit measurements followed by classical feed-forward operations conditioned on the measurement outcome and are an essential building block for quantum error correction. Qubits that are idle during the (long) measurement and classical feed-forward operation are affected by significant coherent $ZZ$ errors. We consider a simple example as a benchmark with a linear chain of three qubits (two data qubits and an auxiliary qubit) and use dynamic circuits to create a Bell state ($1/\sqrt{2}(\ket{00}+\ket{11})$, see Fig.~\ref{fig:dyncir}a. In some experiments, due to complexities of programming the controller, it is be challenging to apply DD during feedforward~\cite{baumer2023efficient}. Therefore, we only consider the application CA-EC to improve the fidelity of preparing this state. Specifically, we first characterize various sources of coherent errors together with the total time of the operation during which the qubits are idling. As we observed earlier, there are coherent $ZZ$ interactions $H_{11}$~\eqref{eq:H11} between neighboring idle qubits. However, when an idle qubit is adjacent to a measured qubit, we can use the information from the measurements to simplify error compensation. Therefore, instead of applying a two-qubit correction to the auxiliary qubit and its data spectator, we append an additional single-qubit $Z$ correction to the conditional, see Fig.~\ref{fig:dyncir}b. The angles in the compensating gates depend on the total time $\tau$ of measurements and feedforward. While the former is known accurately in our experiment ($4~\mu$s), the latter has to be calibrated. Therefore, we vary $\tau$ and measure the fidelity of Bell state preparation.} 

{Without error compensation, the fidelity is $9.5\%$ due to the large accumulation of the $ZZ$ errors, while using CA-EC, the fidelity increases significantly by more than $8\times$ to $78.1\%$ at the optimal feedforward time of $1.15~\mu$s, see Fig.~\ref{fig:dyncir}c. This optimal calibrated  time agrees with independent estimates of the feedforward time using other techniques. }

\subsection{{Combined compiling strategy}}
\begin{figure}[ht]
	\centering
	\includegraphics[width=\columnwidth]{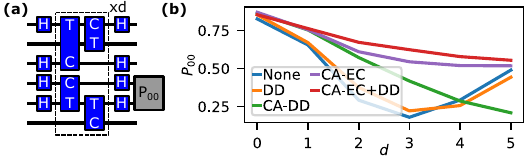}
	\caption{{Combined compiling strategy. (a) A Floquet time-evolution circuit with 6 qubit, with the final measurement of $P_{00}$ on two of the qubits. This quantity should stay at 1 at every time-step $d$.  (b) We apply different suppression strategies and observe that the combined strategy (CA-EC+DD) outperforms its constituent methods applied individually. Experiments are performed on {\em ibm\_penguino1}.}}
	\label{fig:ca-dd-ec}
\end{figure}

{Finally, we consider  combining CA-EC and CA-DD in applications. In general, there are several factors such as the relative strength of coherent $ZZ$ versus incoherent errors such as $T_2$, as well as single qubit gate errors that one needs to consider when applying these schemes together. For example, if errors are dominated by slow incoherent noise, it is more beneficial to use CA-DD, while in cases where the single qubit gates are too noisy or when the idle time is too short to add DD pulses it is beneficial to use CA-EC. Here, we consider a straightforward combination, where we first use CA-DD to suppress possible noise instances and use CA-EC to suppress those that CA-DD was unable to address (e.g., case IV in Fig.~\ref{fig:isolated_cases}). In a 6-qubit experiment with a Floquet-type circuit, we show that the combined method (CA-EC+DD) indeed performs better than its constituent methods, see Fig.~\ref{fig:ca-dd-ec}.       }

\section{Related work}

Dynamical decoupling has a long history in quantum systems~\cite{Hahn1950}, though the vast majority of demonstrations have been in small systems or via manual pulse insertion~\cite{pokharel2018demonstration, Ezzell_DD_2022, niu2022analyzing}. Recently, several works have applied DD with success to larger multi-qubit circuits, but in a uniform fashion across all qubits~\cite{Jurcevic_Demonstration_2021, chen2021exponential}, which fails to address non-local errors. More recently, several experiments have demonstrated crosstalk suppression using staggered dynamical decoupling~\cite{Tripathi2022,zhou2022quantum}, building on theoretical proposals from decades ago~\cite{pawel2006,Leung_Simulation_2002}. However these have been manual with no compiler support. The idea of adapting DD sequences to different parts of the circuit has gained traction recently~\cite{smith2021error, das2021adapt, ravi2022vaqem,tong2024empirical} but these often require tuning the placements via many additional circuit executions and are not informed by the physical properties of the noise. Suppressing crosstalk by changing the circuit schedule or tuning qubit frequencies have also been studied in Ref.~\cite{murali2020software, ding2020systematic}, though with limited experimental evaluation and non-scalable compilation via exact solvers. 
Lastly, there has been some research on adapting circuits to combat coherent errors, though mainly focused on gate errors and not correlated spectator errors~\cite{white2023unifying}. Ref.~\cite{zhang2022hidden} alternates the physical implementation of a gate within a circuit to get an overall suppression. Similarly, Ref.~\cite{lao2022software} re-compiles the circuit after considering errors arising from gates on a flux-tunable architecture, and evaluate the proposal numerically.

\section{Conclusion and outlook}

We presented a characterization of various sources of coherent and correlated errors affecting fixed-frequency superconducting qubits, and proposed two methods for their suppression: (i) context-aware DD, (ii) context-aware error compensation. Our experiments show substantial improvements to execution fidelities of large circuits, with no extra qubits, gates or samples required. In practical terms, these gains translate to orders of magnitude smaller overhead when performing quantum error mitigation or correction. Correlated errors appear on other quantum computing platforms as well, and our proposals can in principle be adapted for their suppression. 

While coherent errors are more detrimental to quantum circuits than stochastic errors, we exploit the unitary nature of these errors and modify the circuit to counter them. The key insight is that these modifications must be informed by the spatial and temporal context of instructions being executed, and a knowledge of the underlying sources of error.

At a broader level, our study has some implications on designing quantum circuits and architectures. For example, some sources of crosstalk, such as those that survive two nearest-neighbor targets, are hard to suppress using either of our approaches. One could therefore ask a compiler to not schedule circuits with these undesirable contexts, or try to suppress these forms of spectator crosstalks during gate calibration.

\section*{acknowledgments}
	The authors would like to thank Gregory Quiroz, Luke Govia, Lev Bishop, David McKay, Blake Johnson, and Jay Gambetta for insightful discussions. 
	This work was partially supported by the U.S. Department of Energy, Office of Science, National Quantum Information Science Research Centers, Co-design Center for Quantum Advantage (C2QA) under contract number DE-SC0012704.
\end{document}